\documentstyle[12pt,graphicx]{article}
\title{On contributions of fundamental particles to the vacuum
energy}
\author{G.E. Volovik
\\Low Temperature Laboratory, Helsinki University of
Technology\\
P.O.Box 2200, FIN-02015 HUT, Finland\\
and\\
  L.D. Landau Institute for
Theoretical Physics\\  Kosygin Str. 2, 117940 Moscow, Russia}
\begin{document}
\maketitle

\abstract{
Recently different regularization schemes for calculations of the vacuum
energy stored in the zero-point motion of fundamental fields were
discussed. We show that the contribution of the fermionic and bosonic
fields to the energy of the vacuum depends on the physical realization of
the vacuum state. The energy density of the homogeneous equilibrium
vacuum is zero irrespective  of the fermionic and bosonic content of the
effective theories in the infra-red corner. The contribution of the
low-energy fermions and bosons becomes important when the coexistence of
different vacua is considered, such as the bubble of the true vacuum
inside  the false one. We consider  the case when these vacua differ only
by the masses of the low-energy fermionic fields,
$M_{\rm true}>M_{\rm false}$, while their ultraviolet structure is
identical. In this geometry the energy density of the false vacuum
outside the bubble is zero, $\rho_{\rm false}=-P_{\rm false}=0$, which
corresponds to zero cosmological constant. The energy density of the true
vacuum inside the bubble is
$\rho_{\rm true}=-P_{\rm true}\propto -\Lambda^2(M^2_{\rm true}- M^2_{\rm
false})$, where
$\Lambda$ is the ultraviolet cut-off.}

%PACS: 47.20.Ma, 67.57.Np, 68.05.$-$n

\vfill\eject

\section{Introduction}

Recently the problem of the contribution of different fermionic and
bosonic fields to the vacuum energy was revived in relation to the
cosmological constant problem (see e.g.
\cite{OssolaSirlin,Akhmedov}). The general form of the vacuum energy
density under discussion was
\begin{equation}
\rho  =a_4 \Lambda^4 + a_2\Lambda^2M^2 + a_0M^4\ln{\Lambda^2\over M^2}~,
\label{General}
\end{equation}
where $M$ is the mass of the corresponding field, and $\Lambda$ is the
ultraviolet energy cut-off. Different regularization schemes were
suggested in order to obtain the dimensionless parameters $a_i$.

Here we consider this problem from the point of view of the effective
relativistic quantum field theory emergent in the low-energy corner of the
quantum vacuum, whose ultraviolet structure is known. It appears that the
parameters $a_i$ depend on the details of the trans-Planckian physics.
But in addition we find that even if the two vacua have completely
identical structure throughout all the scales, the parameters $a_i$
depend on the arrangement and geometry of the vaccum state. In particular,
if the vacuum is completely homogeneous and static, all the parameters
vanish,
$a_4=a_2=a_0=0$, which corresponds to zero cosmological constant. While
in the presence of the interface between two different vacua, the energy
density can depend not only on the mass $M$, but also on the mass of the
quantum field in the neighboring vacuum.

\begin{figure}
  \centerline{\includegraphics[width=0.7\linewidth]{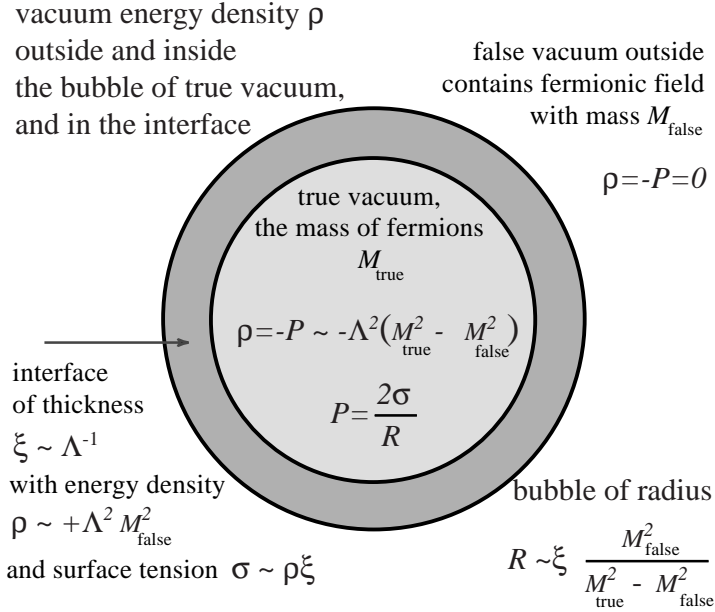}}
   \caption{Critical bubble of true vacuum  inside the false
one. }
   \label{VacuumFig}
\end{figure}

As an example we consider two vacua, whose structure is identical in the
high-energy limit, while in the low-energy
corner they have the same fermionic content, but their low-energy
fermions have slightly different masses.  Let us assume that both vacua
correspond to the local minima, and the route between these vacua lies
only through the vacuum with Weyl fermions of zero mass. Then these vacua
can coexist in the configuration shown in
Fig. \ref{VacuumFig}. The false vacuum occupies the whole Universe
except for the bubble of the true vacuum separated from the false vacuum
by the interface where the vacuum has zero-mass fermions.  Let us consider
the behavior of the vacuum energy density $\rho$ outside and inside the
bubble, when this confiuration is static, i.e. when the bubble has the
critical size $R$ corresponding to the saddle point of energy functional.

The picture, which is supported by the effective relativistic quantum
field theory emerging in condensed matter systems, where the structure
of the vacuum is known both at high and low  energy  (Sec.
\ref{VacuumEnergySec}), is shown in
Fig. \ref{VacuumFig}. The energy density of the false vacuum  outside the
bubble does not contain any fermionic mass $M$, since this energy is
simply zero;
\begin{equation}
  \rho_{\rm false}=-P_{\rm false}=0~.
\label{FalseVacuum}
\end{equation}
  This
corresponds to zero cosmological constant in any homogeneous vacuum, if
the vacuum is stationary, corresponds to the extremum of
the energy functional, and is isolated from the environment
\cite{Book}.

The energy density of the true vacuum inside the bubble,
\begin{equation}
  \rho_{\rm true}=-P_{\rm true}\sim -\Lambda^2\left(M_{\rm
true}^2-M_{\rm false}^2\right)\sqrt{-g}<0~.
\label{TrueVacuum}
\end{equation}
contains the quadratic in
$M_{\rm true}$ term with negative sign, but it also depends quadratically
on the fermionic mass $M_{\rm false}$ in the neighboring vacuum.
On the other hand, if the true vacuum occupies the whole space and
becomes equilibrium, its vacuum energy density is zero again, $\rho_{\rm
true}=-P_{\rm true}=0$.

All this can be obtained  without invoking the
microscopic (ultraviolet) structure of the quantum vacuum, using only the
general arguments of the vacuum stability. However, here we shall use the
microscopic model from which the relativistic quantum field theory with
massive fermions emerges in the low-energy corner.

\section{Vacuum energy in effective quantum field theory}
\label{VacuumEnergySec}

\subsection{Effective quantum field theory with Dirac fermions}

Massive relativistic fermions are realized as the low-energy Bogoliubov
quasiparticles in the class of the spin-triplet
superfluids/superconductors (see section 7.4.9 of \cite{Book}). The
Bogoliubov--Nambu Hamiltonian for fermionic quasiparticles (analog of
elementary particles) is the
$4\times 4$ matrix
\begin{eqnarray}
H_{\bf p}=\check \tau^3 \left({p^2\over 2m}-\mu\right) +
\check\tau^1
\sigma^\mu p_i~  e_ \mu^i~,
\label{BogoliubovNambuHam}
\\
e_ \mu^i={\Lambda\over p_F}\left(\hat x_\mu \hat x^{i}
+  \hat y_\mu \hat y^{i}\right) +{M\over p_F}\hat z_\mu \hat
z^{i}~~,~~{p_F^2\over 2m}=\mu~.
\label{BogoliubovNambuHam2}
\end{eqnarray}
Here  $\mu$ is the chemical potential of the particles (atoms) forming the
liquid (analog of the quantum vacuum); $m$ is their mass; $\check \tau^a$
and
$\sigma^\mu$ are $2\times 2$ Pauli matrices describing the
Bogoliubov-Nambu spin and the ordinary spin of particles correspondingly;
$\Lambda$ and $M$ are amplitudes of the order parameter $e_ \mu^i$.

One should not confuse particles which form
the vacuum (atoms of the liquid) and quasiparticles --
  excitations above the vacuum -- which form the analog
of  matter in quantum liquids and correspond to elementary particles.
Quasiparticles do not scatter
on the atoms of the liquid if the liquid is in its ground state, and thus
for quasiparticles the ground state of the liquid is seen as an
empty space -- the vacuum -- though this space is densly filled by atoms.

The
energy spectrum of quasiparticles in this model  is
\begin{equation}
E^2({\bf  p})=H_{\bf p}^2=\left({p^2\over 2m}-\mu\right)^2
+ \Lambda^2{p_x^2+p_y^2\over p_F^2}+M^2{p_z^2\over p_F^2}~.
\label{InstabilityPlanarPhase}
\end{equation}
The case $M=0$ corresponds to the so-called planar phase, while
$M=\Lambda$ describes the isotropic
B-phase of superfluid $^3$He.

We assume that the interaction of  atoms in the liquid (or electrons in
superconductors), which leads to the formation of the order parameter, is
such that the locally stable vacuum states of the liquid have
$M\ll\Lambda$. Then in the low-energy limit the  Bogoliubov--Nambu
Hamiltonian for quasiparticles transforms to the Dirac Hamiltonian
\begin{eqnarray}
H_{\bf p}\approx  c_\parallel \tilde p_z\check \tau^3 + \check
\tau^1 (c_\perp p_x \sigma_x +c_\perp p_y \sigma_y + M \sigma_z)
~,
\label{BogoliubovNambuHamiltonian}
\\
\tilde p_z=p_z-p_F~~,~~c_\parallel={p_F\over m}~~,~~c_\perp={\Lambda\over
p_F}~,
\label{BogoliubovNambuHamiltonian2}
\end{eqnarray}
with the relativistic spectrum
\begin{eqnarray}
E^2({\bf  p})\approx g^{ik}\tilde p_i\tilde p_k +M^2~,
\label{RelSpectrum}
\\
\tilde {\bf p}={\bf p}-p_F\hat{\bf z}~~,~~
g^{xx}=g^{yy}=c_\perp^2
  ~~,~~  g^{zz}=c_\parallel^2 ~.
\label{RelSpectrum2}
\end{eqnarray}
Here $c_\parallel$ is the `speed of light' propagating along the
$z$-axis, while in transverse direction the `light' propagates
with the speed $c_\perp$. Though in this model the effective speed of
light depends on the direction of propagation, this anisotropy
can be removed by rescaling along the $z$-axis. The speed of light is not
fundamental, since it is determined by the material parameters of the
microscopic system. However, it is fundamental from the point of view of
all inner observers who consist of the low-energy quasiparicles. For them
the speed of light does not depend on direction, and they believe
in the laws of special relativity since these laws can be confirmed by all
experiments (including the Michelson--Morley measurements of the speed of
light) which use clock and rods made of the low-energy quasiparticles. The
high-energy observer will not agree with that: for us the original model
(\ref{BogoliubovNambuHam}) has no Lorentz invariance, and the speed of
light is anisotropic.

After rescaling along the
$z$-axis the quasiparticles become the complete
analog of the Standard Model fermions with mass $M$:   the left-handed and
right-handed chiral quasiparticles of the planar state with $M=0$
are hybridized to form the Dirac particles with the mass $M$.
Note, that together with the effective Dirac fermions, also the effective
gravity and effective $U(1)$ and $SU(2)$ gauge fields (with `photons'
and `gauge bosons') emerge in the low-energy corner of this model, with
all the accompaning phenomena, such as chiral anomaly, running
couplings, etc.
\cite{Book}. The reason for such close analogy is the momentum space
topology of the quantum vacuum, which is common for the ground state of
the considered liquid and for the quantum vacuum of the Standard Model.
The parameter
$\Lambda$ plays the role of the Planck energy scale in the effective
theory.

\subsection{Vacuum energy}

This model can serve for the consideration of the contribution of the
Dirac fermions to the energy density of the  vacuum in equilibrium. In
the many body system (superfluid liquid which contain
many particles) the relevant vacuum energy whose gradient expansion
gives rise to the effective quantum field theory for quasiparticles at
low energy is $\left<{\cal H}-\mu {\cal N}\right>_{\rm eq~vac}$,
\cite{AGDbook} where ${\cal H}$ is the Hamiltonian of the system,
${\cal N}$ is the particle number operator for atoms forming
the liquid, and
$\mu$ is their chemical potential. The energy density of
the superfluid/superconducting ground state (analog of the quantum
vacuum) with the order parameter
$e_
\mu^i$ is given by (see e.g. Eq.(5.38) in \cite{MineevSamokhin})
\begin{equation}
\rho  ={1\over V}\left<{\cal H}-\mu {\cal N}\right>_{\rm
vac}={1\over V}\left<{\cal H}-\mu {\cal N}\right>_{{\rm vac}~e_ \mu^i=0}
-{p_F^3 m \over  12 \pi^2\hbar^3}  e_ \mu^i e_ \mu^i ~,
\label{FermiGasSubtraction1}
\end{equation}
where $V$ is the volume of the system. For the order parameter
(\ref{BogoliubovNambuHam2}) which gives rise to the relativistic fermions
at low energy one has
\begin{equation}
\rho  =\rho(0) - {\sqrt{-g}
\over  12 \pi^2\hbar^3} \left(2\Lambda^4
+\Lambda^2M^2\right)~,
\label{FermiGasSubtraction}
\end{equation}
where $\sqrt{-g}=   c_\parallel^{-1} c_\perp^{-2}$ according to
Eq.(\ref{RelSpectrum2}); and $\rho(0)\equiv \rho(e_ \mu^i=0)$
is the energy density in the normal (non-superfluid) state of the liquid,
in which $e_\mu^i=0$.  The equation
(\ref{FermiGasSubtraction}) demonstrates that the amplitude
$\Lambda$ in the order parameter expression (\ref{BogoliubovNambuHam2})
is the proper ultraviolet cut-off for the estimation of the vacuum energy
related to the effective relativistic fermions, while the contribution
$\rho(0)$, which does not depend on $\Lambda$ and $M$, comes from the more
fundamental physics of the normal state of the liquid at energies well
above the energy scales $\Lambda$ and $M$. In this model the term
$M^4 \ln (\Lambda^2/M^2)$ is absent, though it naturally appears in all
the regularization schemes discussed in \cite{OssolaSirlin,Akhmedov}.

The dimensionless factors in front of $\Lambda^4$ and $\Lambda^2 M^2$ were
obtained using the microscopic model in Eqs. (\ref{BogoliubovNambuHam2})
and (\ref{InstabilityPlanarPhase}). In principle, the low-energy fermions
in Eq.(\ref{RelSpectrum}) can be obtained from different microscopic
theories, and one finds that the dimensionless factors in front of
$\Lambda^4$ and $\Lambda^2 M^2$ depend on the details of the Planck
physics. In other words, they depend on the regularization imposed by the
ultraviolet physics, which cannot be found within the effective
low-energy theory.

It is rather natural to think that the equation
(\ref{FermiGasSubtraction}) reflects the general
structure (\ref{General}) of the vacuum energy in terms of the massive
fields discussed in
\cite{OssolaSirlin,Akhmedov}. Moreover, from the point of view of
the low-energy observers who are made of quasiparticles and live in the
liquid and for whom the effective theory is fundamental, the cosmological
constant in their world must have the natural value of order $\Lambda^4$.
However, this is not so. The vacuum energy contains the contribution
$\rho(0)$ from the vacuum degrees of freedom with energies well above
$\Lambda$ and
$M$.  Together with the sub-Planckian modes these high-energy modes
determine the behavior of the whole vacuum, and in particular the
stability of the static vacuum configurations.  If something happens in
the low-energy corner, so that the vacuum configuration changes, the
high-energy degrees of freedom respond to restore the stability of the
whole vacuum in the new configuration. As a result
$\rho(0)$ is tuned to the low-energy degrees of freedom and thus
becomes dependent on $\Lambda$ and $M$ after the adjustment of the
trans-Planckian degrees to the sub-Planckian ones. This dependence is
different for different realizations of the equilibrium vacuum state.
Moreover, the adjustment can cancel completely or almost completely the
low-energy contributions. Below we consider how this occurs in the
geometry of Fig.
\ref{VacuumFig}.

\subsection{Energy density of the false vacuum}

Let us  assume now that the situation is similar to that of the
Ising ferromagnet in applied small magnetic field which
slightly discriminates between spin-up and spin-down vacua. This means
that there are two vacuum states, false and true, corresponding to two
nearly degenerate local minima whose fermionic masses are almost
the same:
\begin{equation}
0<M^2_{\rm
true}-M^2_{\rm false} \ll M^2_{\rm true} \ll \Lambda^2  ~.
\label{ConditionsForMasses}
\end{equation}
Then we can apply this to the vacuum states in the geometry of Fig.
\ref{VacuumFig}, where the false vacuum is separated by the domain wall
(the interface) from the spherical domain containing the true vacuum.

Let us start with the external domain occupied by the false vacuum. This
domain is open and it occupies the whole space except for the finite
volume. That is why one can apply to this domain the results known for
the infinite homogeneous vacuum.
First, we note that any equilibrium macroscopic system consisting of the
identical elements (atoms in the case of quantum liquids) obeys the
Gibbs-Duhem relation
\begin{equation}
E-\mu N -TS=-PV~,
\label{Gibbs-DuhemRelation}
\end{equation}
where $P$ is the pressure and
$S$ the entropy (see Chapter 10.9 in Ref.  \cite{GibbsDuhemRrlation}).
Applying this to the equilibrium ground state of the system at
$T=0$ (the quantum vacuum), one has for the energy density of the vacuum:
\begin{equation}
\rho  ={1\over V}\left<{\cal H}-\mu {\cal N}\right>_{\rm
vac}={E-\mu N\over V} =-P ~.
\label{Gibbs-Duhem}
\end{equation}
This equation of state, $\rho=-P$, is valid for any homogeneous vacuum
irrespective of whether the system is relativistic or not. If the
effective  relativistic quantum field theory emerges in the low-energy
corner of the system, this $\rho$ becomes the cosmological
cosntant in the low-energy world.

Second, if the
system (the Universe) is isolated from the environment, the external
pressure $P_{\rm external}=0$. Thus for the false vacuum in Fig.
\ref{VacuumFig} one has $P_{\rm false}=P_{\rm external}=0$, which gives
the vanishing energy density of the false vacuum in this geometry:
\begin{equation}
\rho_{\rm false} =-P_{\rm false}=0~.
\label{FalseVacuumEnergy}
\end{equation}
Thus the cosmological constant in the vacuum outside the bubble is zero.
This property does not depend on the low-energy physics, and is
determined by the physics of the whole vacuum including the degrees of
freedom at energies well above the scales $\Lambda$ and $M$. These
high-energy degrees of freedom respond to any change in the low-energy
corner exactly compensating the contribution from the low-energy degrees
of freedom to ensure the zero value of the cosmological
constant in the equilibrium homogeneous vacuum.

\subsection{Energy and pressure of the true vacuum}

Now let us turn to the inner domain occupied by the true
vacuum. If the domain is big enough, $R\gg \Lambda^{-1}$, the vacuum can
be considered as homogeneous, and thus the equation of state
$\rho=-P$ in (\ref{Gibbs-Duhem}) is applicable. However, this vacuum is
not isolated from the environment, and as a result its vacuum energy
density is non-zero. This energy density
$\rho_{\rm true}$ can be found by comparison with the energy density
of the false vacuum using
Eq.(\ref{FermiGasSubtraction}). Since $\rho_{\rm false}=0$ one has
\begin{equation}
\rho_{\rm true} =-P_{\rm true}=\rho_{\rm true}-\rho_{\rm false}= -
{1
\over  12 \pi^2\hbar^3} \sqrt{-g}\Lambda^2\left(M^2_{\rm true}-M^2_{\rm
false}\right)<0~.
\label{FalseTrueDifference}
\end{equation}
In this geometry, the true vacuum has the negative cosmological constant
proportional to the difference of $M^2$ in two vacua. The quartic
term $\propto\Lambda^4$ does not appear in the considered model; it
appears if the two vacua have different physics at energies below
$\Lambda$. Note that if the true vacuum is outside and is in equilibrium,
the cosmological constant in this vacuum will be zero again.

  Let us now estimate the radius $R$ of the static
bubble -- the saddle-point critical bubble -- which is obtained when the
vacuum pressure inside the bubble is compensated by the effect of the
surface tension. Since the external pressure is zero, the pressure
$P_{\rm true}$ within the domain is determined by the surface tension
and the curvature of the domain wall:
\begin{equation}
P_{\rm true} =P_{\rm true} - P_{\rm false}={2\sigma\over R}~.
\label{TrueVacuumPressure}
\end{equation}
Comparing equations (\ref{FalseTrueDifference}) and
(\ref{TrueVacuumPressure}) one obtains the radius of the static
bubble critical bubble. The order of magnitude of the surface
tension is
$\sigma\sim \rho(M=0)\xi$, where $\xi$ is the
thickness of the domain wall; and $\rho(M=0)$ is the
energy density of the vacuum with massless fermions. This energy can be
obtained by comparing it with the energy density of either of the
domains, and it appears to be positive:
\begin{equation}
\rho(M=0)= \rho(M=0)-\rho_{\rm false} =
{\sqrt{-g}
\over  12 \pi^2\hbar^3} \Lambda^2 M^2_{\rm
false}>0~.
\label{InterfaceFalseDifference}
\end{equation}
This gives the following estimation for the radius of the critical bubble:
\begin{equation}
R\sim \xi {M^2_{\rm
false}
\over M^2_{\rm true}-M^2_{\rm
false}} \gg \xi~.
\label{BubbleSize}
\end{equation}
Here $\xi\sim  \hbar c_\parallel/ \Lambda$ in our model, and $\xi\sim
\hbar c/ \Lambda$ in its relativistic counterpart.

\section{Conclusion}
\label{ConclusionSec}

In general, the energy density of the quantum vacuum is determined by
all the vacuum degrees of freedom. They act coherently as the elements
of the same medium, which results in the zero value of the cosmological
constant if the vacuum is equilibrium, homogeneous and is isolated from
the environment.  Thus the contribution of the zero-point motion
of the low-energy fermionic and bosonic fields to the vacuum energy
cannot be singled out from the ultrviolet contributions and thus it cannot
be obtained from the low-energy theory by any regularization scheme.
The low-energy physics is capable to describe only the
contribution of different ifrared perturbations of the vacuum to the
vacuum energy density, such as dilute matter, space curvature, expansion
of the Universe, rotation, Casimir effect, etc.
\cite{Book,Phenomenology}. This gives rise to induced vacuum
energy which is proportional to the
energies of the infrared perturbations, including the energy density
of matter, which is in agreement with observations
\cite{RiessPerlmutter}. If the Universe evolves in time,
cosmological `constant' becomes an evolving
physical quantity, which responds to the combined action of
the evolving perturbations of the vacuum state.

The example which we considered here is somewhat similar to the Casimir
effect, where the difference in energy density between two neghbouring
vacua, in the restricted and unrestricted domains, are calculated using
the low-energy physics. In our case, the two vacua across the interface
(domain wall) also have identical ultraviolet (microscopic) structure and
also have very close low-energy theories, whose fermionic and bosonic
contents are the same, but the masses of the fields are slightly
different. That is why, for the estimation of the difference one can
try to explore the low-energy physics. One can use, for example,  the
regularized equation  for the contribution of the massive quantum field
to the energy-momentum tensor of the vacuum -- the equation (9)  in Ref.
\cite{OssolaSirlin} with minus sign when applied to the
fermionic field:
\begin{equation}
t_{\mu\nu}=-g_{\mu\nu} M^2 \int {d^4p\over
(2\pi)^3}\delta(p^2-M^2)\theta(p_0)~.
\label{Regularization}
\end{equation}
  Then for the difference in energy and pressure between
the true and false vacua one obtains:
\begin{eqnarray}
\rho_{\rm true}-\rho_{\rm false}=P_{\rm false}- P_{\rm true}\\
=
-\left(M_{\rm true}^2-M^2_{\rm false}\right)\int {d^3p\over 2(2\pi)^3}
{1\over cp}  \sim -\sqrt{-g}\Lambda^2\left(M^2_{\rm true}-M^2_{\rm
false}\right) ,
\label{FreeFieldEnergyDifference}
\end{eqnarray}
where $\sqrt{-g}=c^{-3}$. This gives for the energy and pressure
difference the correct dependence on masses,
  the correct sign  and even the correct Gibbs-Duhem
relation $\Delta\rho=-\Delta P$. However, the numerical factor is still
missing because of the quadratic divergence, and this factor depends on
details of the microscopic physics. This demonstrates that the
regularization schemes are not applicable in a strict sense
even when the difference in the vacuum energies is calculated.

As for the energy density itself (the cosmological constant),
it is determined (as in the case of the Einstein and G\"odel Universes
\cite{Phenomenology}) by the equilibrium properties of the whole system
and depends on the geometry.  In particular, for the equilibrium
homogeneous vacuum one obtains
$\rho=-P=0$, irrespective of whether this vacuum is true or false, and
whether it contains massive or massless fermions. On the other hand, the
vacuum inside the bubble is not isolated from the environment, as a
result the energy density of this vaccum is non-zero and is given by the
difference of the quadratic terms in Eq.(\ref{TrueVacuum}). Except for
the numerical factor, this result can be reproduced without consideration
of the microscopic theory. In addition to the general arguments of the
vacuum stability, including the Gibbs-Duhem relation, one can use the
effective  low-energy theory for the order-of-magnitude estimation of the
energy and pressure difference of two neighboring vacua. For our
particular problem, we can use the regularization presented
in Eq.(\ref{Regularization}).

  This work was supported by ESF COSLAB Programme and
by the Russian Foundations for Fundamental Research.


\begin{thebibliography}{15}

\bibitem{OssolaSirlin} G. Ossola and A. Sirlin, Considerations concerning
the contributions of fundamental particles to the vacuum energy
density, hep-ph/0305050.

\bibitem{Akhmedov} E. Kh. Akhmedov, Vacuum energy and relativistic
invariance, hep-th/0204048.

\bibitem{Book} G.E. Volovik, {\it The Universe in a Helium
Droplet}, Clarendon Press,  Oxford (2003).

\bibitem{AGDbook} A.A. Abrikosov, L.P. Gorkov   and I.E. Dzyaloshinskii,
{\it Quantum Field Theoretical Methods in Statistical Physics},
Pergamon, Oxford (1965).

\bibitem{MineevSamokhin} V.P. Mineev and K.V. Samokhin, {\it
Introduction to Unconventional Superconductivity}, Gordon and Breach
Science Publishers (1999).

\bibitem{GibbsDuhemRrlation} B. Yavorsky and A. Detlaf, {\it Handbook of
Physics}, Mir Publishers, Moscow (1975).

\bibitem{Phenomenology} G. E. Volovik,  Phenomenology of
effective gravity,  gr-qc/0304061.

\bibitem{RiessPerlmutter} A. G. Riess, {\it et al.},
Astron. J. {\bf 116}, 1009 (1998); S. Perlmutter,  {\it et al.},
Astrophys. J. {\bf 517}, 565 (1999).

\end{thebibliography}
\end{document}